\documentstyle[amssymb,aps,12pt]{revtex}

\begin{document}
\draft
\author{O. B. Zaslavskii}
\address{Department of Mechanics and Mathematics, Kharkov V.N. Karazin's National\\
University, Svoboda\\
Sq.4, Kharkov 61077, Ukraine\\
E-mail: aptm@kharkov.ua}
\title{Classification of static and homogeneous solutions in exactly solvable
models of two-dimensional dilaton gravity}
\maketitle

\begin{abstract}
We give the full list of types of static (homogeneous) solutions within a
wide family of exactly solvable 2D dilaton gravities with backreaction of
conformal fields. It includes previously known solutions as particular
cases. Several concrete examples are considered for illustration. They
contain a black hole and cosmological horizon in thermal equilibrium,
extremal and ultraextremal horizons, etc. In particular, we demonstrate that
adS and dS geometries can be {\it exact} solutions of semiclassical field
equations for a {\it nonconstant} dilaton field.
\end{abstract}

\pacs{PACS numbers: 04.60.Kz, 98.80.Cq, 11.25.-w}


\section{introduction}

Semiclassical physics of black holes, combining issues of space-time,
thermodynamics and quantum theory is the one of the most fascinating areas
in physics. In our real four-dimensional world high mathematical
complexities obscure the analysis of interplay between these aspects. This
explains why the two-dimensional (2D) black hole physics (and, in more
general settings, 2D dilaton gravity theories) became so popular during last
decade. The powerful incentive was given due to Callan, Giddins, Harvey and
Strominger (CGHS) work \cite{callan} on evaporation of two-dimensional black
holes. Meanwhile, the exact solutions for the metric and dilaton discussed
in \cite{callan} were pure classical. The situation becomes much more
complex if backreaction is taken into account. Then even within the set of
two-dimensional theories it is not a simple task to solve and analyze
semiclassical field equations. As a consequence, self-consistent
generalization of the CGHS theory turned out to be a non-trivial problem. To
this end, a series of particular exactly solvable models were suggested and
analyzed \cite{bil} - \cite{fub}. In \cite{kaz} there has been suggested an
unified approach based on symmetries of the non-linear sigma model to which
the gravitation-dilaton action is related. This enabled to embrace
previously known exactly solvable models within an unified scheme, the
condition of exact solvability representing some relation between
coefficients which enter the form of the action. This condition was
independently refound in \cite{exact} in a more direct way, starting from
the gravitation-dilaton action itself. It turned out that rather wide
classes of solutions in such theories shares common properties
(thermodynamics, space-time structure, etc.) which were discussed in \cite
{exact}, \cite{thr}. Meanwhile, these classes of solutions do not exhaust
all the possibilities and, in some respect, are not applicable in some
physically interesting situations. For instance, black holes considered in 
\cite{exact}, are always non-extreme. The solutions for the extreme case can
be obtained explicitly on the pure classical level (see, e.g., the recent
paper \cite{gukov}), but this problem becomes much more complex, when
quantum backreaction is taken into account. For exactly solvable models of
2D dilaton gravity extremality can be achieved by special ''tuning''
asymptotic behavior of some action coefficients near the horizon \cite{ext}, 
\cite{mod} for rather special families of solutions in which quantum
stresses diverge on the horizon, the geometry remaining regular there. Being
interesting on its own, such kinds of solutions do not represent, however,
zero temperature black holes in the Hartle-Hawking state. Meanwhile, it was
shown recently \cite{extr2} that the latter type of solutions does appear in
the exactly solvable models but only in some degenerate cases.

Thus, different classes of the same exactly solvable models may exhibit
quite different properties and this motivates constructing the general
scheme which would include all kinds of solutions. This is just the main
purpose of our work. Such classification is a necessary step for better
understanding the structure of exactly solvable models in dilaton gravity.
It may also serve as a basis for diverse set of physical applications, that
are contained in 2D dilaton gravity (see, e.g., recent reviews \cite{dv}, 
\cite{od}).

\section{basic equations}

Hereafter we restrict ourselves to semiclassical dilaton gravity with
backreaction of conformal fields only\footnote{%
We do not consider additional scalar, Yang-Mills or fermion fields \cite
{fil1} - \cite{pelzer}, theories nonlinear with respect to curvature \cite
{eliz4}, etc, where, however, exact integrability is achieved for the
classical case only.}. Consider the action 
\begin{equation}
I=I_{0}+I_{PL}\text{,}  \label{action}
\end{equation}
where 
\begin{equation}
I_{0}=\frac{1}{2\pi }\int_{M}d^{2}x\sqrt{-g}[F(\phi )R+V(\phi )(\nabla \phi
)^{2}+U(\phi )]  \label{clac}
\end{equation}
and the Polyakov-Liouville action \cite{pl} 
\begin{equation}
I_{PL}=-\frac{\kappa }{2\pi }\int_{M}d^{2}x\sqrt{-g}[\frac{(\nabla \psi )^{2}%
}{2}+\psi R]  \label{PL}
\end{equation}
is responsible for backreaction. Here the function $\psi $ obeys the
equation 
\begin{equation}
\Box \psi =R\text{,}  \label{psai}
\end{equation}
where $\Box =\nabla _{\mu }\nabla ^{\mu }$, $\kappa =N/24$ is the quantum
coupling parameter, $N$ is number of scalar massless fields, $R$ is a
Riemann curvature. We omit the boundary terms in the action as we are
interested only in field equations and their solutions.

From eqs. (\ref{action}) - (\ref{psai}) one can infer field equations (see
below) which are valid for any gravitation-dilaton system of the kind under
discussion. Meanwhile, our main goal is to analyze possible exactly solvable
cases. The typical representative of the corresponding family reads
\begin{equation}
F=\exp (-2\phi )+2\kappa (d-1)\phi \text{, }V=4\exp (-2\phi )+2(1-2d)\kappa
+4C(e^{-2\phi }-\kappa d)^{2}\text{, }U=4\lambda ^{2}\exp (-2\phi )\text{,}
\label{cn}
\end{equation}
$\lambda $ and $d$ are constants. If $C=0$, this model turns to that
suggested in \cite{cruz}. In turn, it includes different particular known
models. For example, in the case $d=0$ one obtains the model suggested in 
\cite{bose}, if $d=1/2$ it coincides with the RST\ model \cite{rst}.
Meanwhile, the family of exactly solvable models under discussion in our
paper is wider than (\ref{cn}), including it only as a particular class.

Let us return to the issue of field equations in the generic case. Varying
the action with respect to a metric gives us $(T_{\mu \nu }=2\frac{\delta I}{%
\delta g^{\mu \nu }})$: 
\begin{equation}
T_{\mu \nu }\equiv T_{\mu \nu }^{(0)}-T_{\mu \nu }^{(PL)}=0\text{,}
\label{6}
\end{equation}
where 
\begin{equation}
T_{\mu \nu }^{(0)}=\frac{1}{2\pi }\{2(g_{\mu \nu }\Box F-\nabla _{\mu
}\nabla _{\nu }F)-Ug_{\mu \nu }+2V\nabla _{\mu }\phi \nabla _{\nu }\phi
-g_{\mu \nu }V(\nabla \phi )^{2}\}\text{,}  \label{7}
\end{equation}
\begin{equation}
T_{\mu \nu }^{(PL)}=\frac{\kappa }{2\pi }\{\partial _{\mu }\psi \partial
_{\nu }\psi -2\nabla _{\mu }\nabla _{\nu }\psi +g_{\mu \nu }[2R-\frac{1}{2}%
(\nabla \psi )^{2}]\}  \label{8}
\end{equation}

Variation of the action with respect to $\phi $ gives rise to the equation 
\begin{equation}
R\frac{dF}{d\phi }+\frac{dU}{d\phi }=2V\Box \phi +\frac{dV}{d\phi }(\nabla
\phi )^{2}\text{.}  \label{9}
\end{equation}

In general, field equations cannot be solved exactly and the function $\psi $%
, the dilaton $\phi $ and metric depend on both time-like ($t$) and
space-like ($\sigma $) coordinates: $\psi =\psi (t$,$\sigma )$, $\phi =\phi
(t$, $\sigma )$. In what follows we restrict ourselves to such kind of
solutions that $\psi $ can be expressed in terms of $\phi $ only: $\psi
=\psi (\phi )$. We will see that this leads to the existence of the Killing
vector. On the other hand, as all static or homogeneous solutions depend on
one variable, one may exclude it and express $\psi $ in terms of $\phi $.
Thus, the assumption $\psi =\psi (\phi )$ turns out to be equivalent to the
static or homogeneous character of solutions.

Let us take the trace of eqs. (\ref{6})-(\ref{8}) and eq. (\ref{9})$.$
Denoting 
\begin{equation}
\tilde{F}\equiv F-\kappa \psi ,U\equiv \Lambda e^{\int d\phi \omega }\text{,}
\label{fu}
\end{equation}
we get 
\begin{equation}
U=\Box \tilde{F}  \label{12}
\end{equation}
\begin{eqnarray}
&&A_{1}\Box \phi +A_{2}(\nabla \phi )^{2}=0\text{,}  \nonumber \\
&&A_{1}=(u-\kappa \omega )\psi ^{\prime }+\omega u-2V\text{,}  \label{13} \\
&&A_{2}=(u-\kappa \omega )\psi ^{\prime \prime }+\omega u^{\prime
}-V^{\prime }\text{,}  \nonumber
\end{eqnarray}
where $u\equiv F^{\prime }$ and prime throughout the paper denotes
differentiation with respect to $\phi .$ For arbitrary coefficients $%
A_{1}(\phi )$, $A_{2}(\phi )$ eq.(\ref{13}) cannot be solved exactly. This
can be done, however, under some restrictions on the form of the
coefficients $A_{1}$, $A_{2}$. Let us demand that 
\begin{equation}
A_{1}=(u-\kappa \omega )\chi ^{^{\prime }}\text{, }A_{2}=(u-\kappa \omega
)\chi ^{^{\prime \prime }}  \label{a}
\end{equation}
where $\chi =\chi (\phi )$ and $\Box \chi =0$. Then it follows that $\psi
=\psi _{0}+\chi $, where 
\begin{equation}
\psi _{0}^{\prime }=\frac{2V-\omega u}{u-\kappa \omega }\text{,}  \label{14}
\end{equation}
which enables us to find at once $\psi _{0}$ in terms of known functions $u$%
, $V$, $\omega $ by direct integration. Demanding that both equations in (%
\ref{a}) be consistent with each other, we obtain the restriction on the
action coefficients 
\begin{equation}
u^{\prime }(2V-\omega u)+u(u\omega ^{\prime }-V^{\prime })+\kappa (\omega
V^{\prime }-2V\omega ^{\prime })=0  \label{15}
\end{equation}

This equation can be solved: 
\begin{equation}
V=\omega (u-\frac{\kappa \omega }{2})+C(u-\kappa \omega )^{2}\text{,}
\label{v}
\end{equation}
where $C$ is a constant.

The fact that the function $\psi $ is defined up to the function whose
Laplacian vanishes is explained by eq. (\ref{psai}) which is, in fact, is
the definition of $\psi $. The presence of $\chi $ reveals itself in the
nature of quantum state (see below). Eq. (\ref{v}) is just the condition
obtained in \cite{exact}, so account for $\chi $ does not generate new types
of exactly solvable models but extends the set of solutions within these
models.

With eq. (\ref{12}) taken into account, the field equations (\ref{6}) - (\ref
{8}) can be rewritten in the form 
\begin{equation}
\lbrack \xi _{1}\Box \phi +\xi _{2}(\nabla \phi )^{2}]g_{\mu \nu }=2(\xi
_{1}\nabla _{\mu }\nabla _{\nu }\phi +\xi _{2}\nabla _{\mu }\phi \nabla
_{\nu }\phi )  \label{16}
\end{equation}
where $\xi _{1}=\frac{d\tilde{F}}{d\phi }$, $\xi _{2}=\frac{d^{2}\tilde{F}}{%
d\phi ^{2}}-\tilde{V}$, $\tilde{V}=V-\frac{\kappa }{2}(\frac{d\psi }{d\phi }%
)^{2}$. Let us multiply this equation by the factor $\zeta $ chosen in such
a way that $\xi _{2}\zeta =\frac{d(\xi _{1}\zeta )}{d\phi }$. Then eq. (\ref
{16}) turns into 
\begin{equation}
g_{\mu \nu }\Box \mu =2\nabla _{\mu }\nabla _{\nu }\mu \text{,}  \label{17}
\end{equation}
where by definition $\mu ^{\prime }=\xi _{1}\zeta .$ This equation takes the
same form as eq.(2.24) from \cite{solod} and entails the same general
conclusion about the existence of the Killing vector $l_{\alpha
}=\varepsilon _{\alpha }^{\beta }\mu _{,\beta }$. In the present paper we
consider the case when the Killing vector is time-like everywhere that gives
rise to static solutions and mainly concentrate on black hole ones.

It is convenient to work in the conformal gauge 
\begin{equation}
ds^{2}=g(-dt^{2}+d\sigma ^{2})\text{,}  \label{g}
\end{equation}
where, in accordance with the choice of the Killing vector, $g=g(\sigma )$
and does not depend on a time-like coordinate $\sigma $. In the gauge (\ref
{g}) the curvature 
\begin{equation}
R=-g^{-1}\frac{\partial ^{2}\ln g}{\partial \sigma ^{2}}.  \label{R}
\end{equation}
Eq. (\ref{12}) takes the form 
\begin{equation}
\Lambda e^{\eta }=\frac{\partial ^{2}\tilde{F}}{\partial \sigma ^{2}}g^{-1}%
\text{, }\eta =\int d\phi \omega \text{.}  \label{cons}
\end{equation}

Now for any function $f(\sigma )$ we have $\Box f=g^{-1}\frac{\partial ^{2}f%
}{\partial ^{2}\sigma }$ whence it is clear that $\chi =\gamma \sigma $,
where $\gamma $ is a constant. Thus, we have 
\begin{equation}
\psi =\psi _{0}+\gamma \sigma \text{,}  \label{psg}
\end{equation}
where $\psi _{0}$ is defined according to (\ref{14}). It follows from (\ref
{psai}) that 
\begin{equation}
g=e^{-\psi -a\sigma }=e^{-\psi _{0}-\delta \sigma }\text{,}
\end{equation}
where $a$ is a constant, $\delta =\gamma +a$. After simple rearrangement the
(00) and (11) field equations (\ref{6}), (\ref{16}) with the metric in the
conformal gauge (\ref{g}) are reduced to one equation 
\begin{equation}
\xi _{1}\frac{d^{2}\phi }{d\sigma ^{2}}+\xi _{2}\left( \frac{d\phi }{d\sigma 
}\right) ^{2}-\xi _{1}g^{-1}\frac{dg}{d\phi }\frac{d\phi }{d\sigma }=0\text{.%
}  \label{x}
\end{equation}

It is convenient to split coefficients in eq. (\ref{x}) into two parts
singling out the term which is built up with the help of $\psi _{0}$: $\xi
_{1}=\xi _{1}^{(0)}-\kappa \gamma \frac{d\sigma }{d\phi }$, $\xi _{2}=\xi
_{2}^{(0)}-\kappa \gamma \frac{d^{2}\sigma }{d\phi ^{2}}+\kappa [\frac{d\eta 
}{d\phi }\gamma \frac{d\sigma }{d\phi }+\frac{1}{2}(\gamma \frac{d\sigma }{%
d\phi })^{2}]$, 
\begin{equation}
\xi _{1}^{(0)}=\frac{d\tilde{F}^{(0)}}{d\phi },\xi _{2}^{(0)}=\frac{d^{2}%
\tilde{F}^{(0)}}{d\phi ^{2}}-\tilde{V}^{(0)},\tilde{F}^{(0)}=F-\kappa \psi
_{0},\tilde{V}^{(0)}=V-\frac{\kappa }{2}(\frac{d\psi _{0}}{d\phi })^{2}.
\label{0}
\end{equation}

Then eq. (\ref{x}) takes the form 
\begin{equation}
\xi _{1}^{(0)}\frac{d^{2}\phi }{d\sigma ^{2}}+\xi _{2}^{(0)}\left( \frac{%
d\phi }{d\sigma }\right) ^{2}+\xi _{1}^{(0)}\frac{d\phi }{d\sigma }(\frac{%
d\psi _{0}}{d\sigma }+\delta )=\kappa \gamma (\delta -\frac{\gamma }{2})%
\text{.}  \label{f2}
\end{equation}
Let us multiply this equation by the factor $s$ such that $\xi _{2}^{(0)}s=%
\frac{d(\xi _{1}^{(0)}s)}{d\phi }$, $s=\exp [\frac{(\xi _{2}^{(0)}-\xi
_{1}^{(0)^{\prime }})}{\xi _{1}^{(0)}}]=\exp [-\tilde{V}^{(0)}/\xi
_{1}^{(0)}]$.

Then eq. (\ref{f2}) can be cast into the form 
\begin{equation}
\frac{dz}{d\sigma }+z(\frac{d\psi _{0}}{d\sigma }+\delta )=\kappa \gamma
(\delta -\frac{\gamma }{2})s\text{,}  \label{zet}
\end{equation}
where $z=s\xi _{1}^{(0)}\frac{d\phi }{d\sigma }=s\frac{d\tilde{F}^{(0)}}{%
d\sigma }$. It follows from (\ref{14}) and (\ref{v}) that 
\begin{equation}
\psi _{0}=\eta +2CH\text{,}  \label{psi0}
\end{equation}
\begin{equation}
g=e^{-\eta -2CH-\delta \sigma }  \label{gc}
\end{equation}
and 
\begin{equation}
\tilde{F}^{(0)}=H(1-2\kappa C)\text{,}  \label{fh}
\end{equation}
where $H=F-\kappa \eta $. If $\Lambda \neq 0$, the metric function (up to
the constant factor) is equal to 
\begin{equation}
g=\frac{e^{-\delta \sigma }}{U}\text{,}
\end{equation}
\begin{equation}
H=F-\kappa \ln U+const\text{.}
\end{equation}

We obtain from (\ref{v}), (\ref{0}), (\ref{psi0}), (\ref{fh}) 
\begin{equation}
\tilde{V}^{(0)}=(1-2\kappa C)H^{\prime }(\omega +CH^{\prime })\text{, }\xi
_{1}^{(0)}=(1-2\kappa C)H^{\prime }\text{,}  \label{v0}
\end{equation}
whence

\begin{equation}
s=e^{-\eta -CH}\text{.}  \label{dz}
\end{equation}
Then after simple rearrangement eq. (\ref{zet}) gives rise to 
\begin{equation}
\frac{d^{2}H}{d\sigma ^{2}}+C\left( \frac{dH}{d\sigma }\right) ^{2}+\delta 
\frac{dH}{d\sigma }=\alpha \text{,}  \label{eqh}
\end{equation}
where 
\begin{equation}
\alpha =\kappa \gamma (\delta -\frac{\gamma }{2})/(1-2\kappa C).
\label{alpha}
\end{equation}

It is convenient to introduce a new variable $\rho $, where $\left| \rho
\right| =e^{CH}$. Then we have the linear equation 
\begin{equation}
\frac{d^{2}\rho }{d\sigma ^{2}}+\delta \frac{d\rho }{d\sigma }=\alpha C\rho
\label{ro}
\end{equation}

One can seek a solution in the form $\rho \sim e^{\beta \sigma }$, whence we
obtain 
\begin{equation}
\beta ^{2}+\delta \beta -\alpha C=0  \label{beta}
\end{equation}
This equation is quadratic and has two roots $\beta _{1}$, $\beta _{2}$.
Depending on their properties, one can classify all possible types of
solutions and describe their properties.. In a natural way, the solutions
fall into three different classes: I (both $\beta _{1}$, $\beta _{2}$ are
real, $\beta _{1}\neq $ $\beta _{2}$); II ($\beta _{1}$, $\beta _{2}$ are
real, $\beta _{1}=\beta _{2}$), III (roots are complex, $\beta _{1}=$ $\beta
_{2}^{*}$). We describe the results below.

\section{general case, types of solutions}

It is convenient to cast the solutions of eq. (\ref{eqh}) into uniform
formulas: 
\begin{equation}
CH=CH_{0}-\frac{\delta \sigma }{2}+\ln \left| f\right| \text{,}  \label{hf}
\end{equation}
where the function obeys the equation

\begin{equation}
\frac{d^{2}f}{d\sigma ^{2}}=f\varepsilon ^{2}\text{, }\varepsilon ^{2}=\frac{%
\delta ^{2}}{4}+\alpha C\text{.}
\end{equation}
We get the following different cases.

I$_{a}$. $\varepsilon ^{2}>0$, $f=\frac{sh\varepsilon \sigma }{\varepsilon }$%
; I$_{b}$: $f=\frac{ch\varepsilon \sigma }{\varepsilon }$;

II$_{a}$: $\varepsilon =0$, $f=\sigma $; II$_{b}$: $f=1;$

III: $\varepsilon ^{2}\equiv -\varkappa ^{2}<0$, $f=\frac{\sin \varkappa
\sigma }{\varkappa }$.

It follows from (\ref{cons}) that 
\begin{equation}
\frac{\Lambda C}{1-2\kappa C}=e^{2CH_{0}}z\text{,}
\end{equation}
where $z=1$ for the I$_{b}$ case, $z=0=\Lambda $ for II$_{b}$ and $z=-1$ in
cases I$_{a}$, II$_{a}$, III.

The Riemann curvature reads the following.

I$_{a}$, II$_{a}$, III: 
\begin{equation}
R=\frac{UC}{1-2\kappa C}[2+\frac{\omega }{CH^{\prime }}-\frac{1}{%
C^{2}H^{\prime }}\left( \frac{\omega }{H^{\prime }}\right) ^{\prime }q^{2}]%
\text{.}  \label{gcur}
\end{equation}
I$_{b}:$ 
\begin{equation}
R=\frac{UC}{1-2\kappa C}[2+\frac{\omega }{CH^{\prime }}+\frac{1}{%
C^{2}H^{\prime }}\left( \frac{\omega }{H^{\prime }}\right) ^{\prime }q^{2}]
\label{r2}
\end{equation}
II$_{b}$: 
\begin{equation}
R=\frac{e^{\eta +2CH_{0}}}{1-2\kappa C}\frac{1}{C^{2}H^{\prime }}\left( 
\frac{\omega }{H^{\prime }}\right) ^{\prime }\frac{\delta ^{2}}{4}\text{.}
\label{r3}
\end{equation}

Here $q=(\frac{df}{d\sigma }-\frac{\delta }{2}f)$.

In a similar way, we get the general structure of the expression for quantum
stresses. Two nonzero components of quantum stresses are connected for
conformal fields by the well known relationship $T_{0}^{0(PL)}+T_{1}^{1(PL)}=%
\frac{\kappa R}{\pi }$ (see eq. (\ref{8})). Here we list the component $%
T_{1}^{1(PL)}$only. One obtains from (\ref{8}), (\ref{psi0}), (\ref{gc}), (%
\ref{alpha}): 
\begin{eqnarray}
T_{1}^{1} &=&-\frac{1}{4\pi g}[\kappa (\frac{\partial \psi _{0}}{\partial
\sigma }+2\delta )\frac{\partial \psi _{0}}{\partial \sigma }+2\alpha
(1-2\kappa C)]\text{,}  \label{t00a} \\
\frac{\partial \psi _{0}}{\partial \sigma } &=&(\frac{\omega }{CH_{\phi
}^{^{\prime }}}+2)\frac{q}{f}\text{,}  \nonumber
\end{eqnarray}
whence 
\begin{equation}
T_{1}^{1(PL)}=-\frac{\kappa }{4\pi }\frac{\left| UC\right| }{1-2\kappa C}Z%
\text{,}  \label{t00}
\end{equation}

\begin{equation}
Z=(\frac{q\omega }{CH^{^{\prime }}}+2f^{^{\prime }})^{2}-(\delta -\gamma
)^{2}f^{2}\text{,}  \label{z}
\end{equation}
except the case II$_{b}$, when 
\begin{equation}
T_{1}^{1(PL)}=-\frac{\kappa }{4\pi }e^{2CH_{0}+\eta }Z\text{,}
\end{equation}
$Z=\frac{\delta ^{2}}{4}\left( \frac{\omega }{CH^{^{\prime }}}\right)
^{2}-(\delta -\gamma )^{2}$.

\section{Particular cases and limiting transitions}

The solutions obtained depend on several parameters. In what follows it is
assumed that the dilaton is not identically constant. The quantities $%
\Lambda $ and $C$ enter the definition of the action coefficients: $\Lambda $
is the ''amplitude'' of the potential $U$ of a generic model according to
eq. (\ref{fu}), while the parameter $C$ defines the coefficient $V$ of an
exactly solvable one (\ref{v}). Meanwhile, the quantities $\delta $ and $%
\alpha $ are the parameters of the solutions of field equations, they do not
enter the action but characterize the different solutions for the same
model. Let us denote the symbolically [$C$, $\Lambda $]( $\delta $, $\alpha $%
) the solutions with given parameters for a given action, where it is
supposed that the values of parameters differ from zero, unless stated
explicitly. Different limiting cases can be described on the basis of eqs. (%
\ref{eqh})-(\ref{beta}) and eq. (\ref{cons}). Consider first the case

\subsection{$C=0$}

Now 
\begin{equation}
\tilde{F}^{(0)}=H,\psi _{0}=\eta \text{, }g=e^{-\eta -\delta \sigma }\text{.}
\label{c=0}
\end{equation}

\subsubsection{Case [$0$, $0$]}

Then it follows from (\ref{cons}) that $\frac{d^{2}H}{d\sigma ^{2}}=0$.

In the cases [$0$, $0$]($0$, $\alpha $) and [$0$, $0$]($\delta $, $0$)
equations (\ref{eqh}) and (\ref{cons}) are mutually inconsistent, so these
cases cannot be realized.

[$0$, $0]$($0$, $0$) 
\begin{eqnarray}
H &=&A\sigma \text{, }g=e^{-\eta }\text{, }R=\frac{A^{2}}{H^{\prime }}\left( 
\frac{\omega }{H^{\prime }}\right) ^{\prime }e^{\eta }\text{,}  \label{000}
\\
\text{ }T_{1}^{1(PL)} &=&-\frac{\kappa }{4\pi }A^{2}\frac{\omega ^{2}e^{\eta
}}{H^{^{\prime }2}}<0\text{.}  \nonumber
\end{eqnarray}
Here $A$ is an arbitrary constant.

[$0$, $0$]($\delta $, $\alpha $) 
\begin{eqnarray}
&&H=\frac{\alpha }{\delta }\sigma \text{, }R=\frac{\alpha ^{2}}{\delta ^{2}}%
\frac{1}{H^{\prime }}\left( \frac{\omega }{H^{\prime }}\right) ^{\prime
}\exp (\eta +\frac{\delta ^{2}}{\alpha }H)\text{,}  \label{0011} \\
&&T_{1}^{1(PL)}=-\frac{\exp (\eta +\frac{\delta ^{2}H}{\alpha })}{4\pi }%
\alpha [\kappa (\frac{\alpha \omega }{\delta H^{^{\prime }}}+2\delta )\frac{%
\omega }{\delta H^{^{\prime }}}+2]\text{.}  \nonumber
\end{eqnarray}

If $\alpha =A\delta $ and $\delta \rightarrow 0$, while $A$ is kept fixed, (%
\ref{0011}) turns into (\ref{000}).

\subsubsection{Case [$0$, $\Lambda $]}

[$0$, $\Lambda $] ($\delta $, $\alpha $) 
\begin{eqnarray}
&&H=H_{0}+\frac{\alpha }{\delta }\sigma +De^{-\delta \sigma }\text{, }R=%
\frac{e^{\eta }}{H^{^{\prime }}}[\omega \Lambda +\left( \frac{\omega }{%
H_{\phi }^{\prime }}\right) _{\phi }^{\prime }(\frac{\alpha ^{2}}{\delta ^{2}%
}e^{\delta \sigma }-2\alpha D+D^{2}\delta ^{2}e^{-\delta \sigma })]\text{, }
\label{0111} \\
&&T_{1}^{1(PL)}=-\frac{1}{4\pi }e^{\eta +\delta \sigma }\{2\alpha +\kappa (%
\frac{\alpha }{\delta }-D\delta e^{-\delta \sigma })\omega H^{^{\prime
}-1}[2\delta +(\frac{\alpha }{\delta }-D\delta e^{-\delta \sigma })\omega
H^{^{\prime }-1}]\}\text{,}  \nonumber
\end{eqnarray}
\begin{equation}
D\delta ^{2}=\Lambda .  \label{qdl}
\end{equation}

[$0$, $\Lambda $]($\delta $, $0$) 
\begin{eqnarray}
H &=&H_{0}+De^{-\delta \sigma }\text{, }g=e^{-\eta }(H-H_{0})D^{-1}\text{, }%
R=\frac{U}{H^{^{\prime }}}[\omega +(H-H_{0})\left( \frac{\omega }{H^{\prime }%
}\right) ^{\prime }]\text{,}  \label{0110} \\
T_{1}^{1(PL)} &=&\frac{\kappa \omega U}{4\pi H^{^{\prime }}}[2-\frac{\omega
(H-H_{0})}{H^{^{\prime }}}]\text{.}  \nonumber
\end{eqnarray}

It is seen from (\ref{eqh}) and (\ref{cons}) that the solution $[0$, $%
\Lambda ]$ $(0$, $0$) is impossible.

[$0$, $\Lambda $]($0$, $\alpha $) 
\begin{eqnarray}
&&H=\frac{\alpha \sigma ^{2}}{2}+H_{0}\text{, }g=e^{-\eta }\text{, }\Lambda
=\alpha =-\frac{\kappa \gamma ^{2}}{2}<0\text{, }  \label{0101} \\
&&R=U[\frac{\omega }{H^{\prime }}+2\left( \frac{\omega }{H^{\prime }}\right)
^{\prime }\frac{(H-H_{0})}{H^{\prime }}]\text{,}  \nonumber \\
&&T_{1}^{1(PL)}=-\frac{U}{2\pi }[1+\left( \frac{\omega }{H^{\prime }}\right)
^{2}\kappa (H-H_{0})]\text{.}  \nonumber
\end{eqnarray}

The above formulae exhaust all the possibilities for $C=0$.

\subsection{$C\neq 0$}

Let now $C\neq 0$.

\subsubsection{ Case [$C$, $\Lambda $]}

[$C$, $\Lambda $]($0$, $0$). Then it follows directly from eq. (\ref{eqh})
and eq. (\ref{12}) that 
\begin{eqnarray}
H &=&C^{-1}\ln \left| \frac{\sigma }{\sigma _{0}}\right| \text{, }g=e^{-\eta
}(\frac{\sigma _{0}}{\sigma })^{2}\text{,}  \label{h00} \\
\text{ }R &=&\frac{U}{1-2\kappa C}[\frac{\omega }{H^{\prime }}+2C-\left( 
\frac{\omega }{H^{\prime }}\right) ^{\prime }\frac{1}{H^{\prime }C}]\text{,}
\nonumber \\
\Lambda C &=&-(1-2\kappa C)\sigma _{0}^{-2}<0\text{,}  \nonumber \\
\text{ }T_{1}^{1(PL)} &=&\frac{\kappa UC}{4\pi (1-2\kappa C)}(2+\frac{\omega 
}{CH^{^{\prime }}})^{2}<0\text{,}  \nonumber
\end{eqnarray}
where $\sigma _{0}$ is a constant.

[$C$, $\Lambda $]($\delta $, $0$): this case can be obtained by putting $%
\alpha =0$ directly in the formulas for the case I. However, we list
equations explicitly since this value is singled out from the physical
viewpoint, giving a typical black hole in the Hartle-Hawking state \cite
{exact}. Here $D$ is a constant, the factor $C$ is singled out for
convenience. 
\begin{eqnarray}
&&e^{CH}=e^{CH_{0}}\left| 1+DCe^{-\delta \sigma }\right| \text{, }g=e^{-\eta
}\frac{[e^{C(H-H_{0})}\nu -1]}{DC}e^{-2CH}\text{,}  \label{1110} \\
&&R=\frac{U}{(1-2\kappa C)}\{2C+\frac{\omega }{H^{\prime }}+\frac{1}{%
CH^{\prime }}\left( \frac{\omega }{H^{\prime }}\right) ^{\prime }[\nu
e^{C(H-H_{0})}-1]\}\text{,}  \nonumber \\
&&\Lambda =e^{2CH_{0}}(1-2\kappa C)D\delta ^{2}\text{, }  \nonumber \\
&&T_{1}^{1(PL)}=\frac{UC}{4\pi }\frac{\kappa }{1-2\kappa C}(\frac{\omega }{%
CH^{\prime }}+2)\{2+\frac{\omega }{CH^{\prime }}[1-\nu e^{C(H-H_{0})}]\}%
\text{,}  \nonumber
\end{eqnarray}
$\nu $=sign$(1+DCe^{-\delta \sigma })$. If $\nu >0$, our solution, written
in the conformal gauge, corresponds to eq. (25) of Ref. \cite{exact}, where
the Schwarzschild gauge was used.

[$C$, $\Lambda $]($0$, $\alpha $)$.$ It can be obtained directly from types
I or III by putting $\delta =0$.

\subsubsection{Case [C, 0]}

[$C$, $0$]($\delta $, $\alpha $): 
\begin{eqnarray}
&&CH=CH_{0}+\beta _{\pm }\sigma \text{, }g=\exp (-\eta -2CH_{0}\mp
2\varepsilon CH\beta _{\pm }^{-1})\text{, }R=\frac{\beta _{\pm }^{2}}{%
C^{2}H^{\prime }}\left( \frac{\omega }{H^{\prime }}\right) ^{\prime }e^{\eta
+2CH_{0}\pm 2\varepsilon CH\beta _{\pm }^{-1}}\text{,}  \label{1011} \\
&&T_{1}^{1(PL)}=-\frac{1}{4\pi }\exp (\eta +2CH_{0}\pm 2\varepsilon \frac{CH%
}{\beta _{\pm }})\{\kappa \beta _{\pm }(2+\frac{\omega }{CH^{^{\prime }}}%
)[2(\beta _{\pm }+\delta )+\frac{\beta _{\pm }\omega }{CH^{^{\prime }}}%
)]+2\alpha (1-2\kappa C)\}\text{,}  \nonumber
\end{eqnarray}
$\beta _{\pm }$ are the roots of eq. (\ref{beta}). The solution [$C$, $0$]($%
0 $, $\alpha $) does not bring any qualitative new features and can be
obtained directly form (\ref{1011}) by putting $\delta =0$.

[$C,0$]($\delta _{0}$,$\alpha $): 
\begin{equation}
H=H_{0}-\frac{\delta \sigma }{2C}\text{, }g=e^{-\eta -2CH_{0}}\text{,}
\end{equation}
\begin{equation}
R=-\frac{\alpha }{CH^{\prime }}\left( \frac{\omega }{H^{\prime }}\right)
^{\prime }e^{\eta +2CH_{0}}\text{,}
\end{equation}

\begin{equation}
T_{1}^{1(PL)}=\frac{\alpha }{4\pi }(\frac{\kappa \omega ^{2}}{CH^{\prime 2}}%
-2)e^{\eta +2CH_{0}}\text{.}
\end{equation}
The solution [$C$, $0$]($\delta $, $0$) can be obtained from (\ref{1011}) by
putting $\alpha =0$.

The solution [$C$, $0$]($0$, $0$) does not exist.

The solutions with $C=0$ and $C\neq 0$ are described by qualitatively
different formulas, so one may ask in what way the first class can be
obtained from the second one by limiting transition $C\rightarrow 0$. In
such a transition one should carefully take into account not only terms with 
$C=0$ but also terms linear in $C$ and, if necessary, make a shift in the
coordinate. For example, compare the cases I and $[0,\Lambda ](\delta
,\alpha )$. In the formulas for roots of eq. (\ref{hf}) we have in the limit
under consideration (let for definiteness $\delta >0$): $\varepsilon =\frac{%
\delta }{2}+\frac{\alpha C}{\delta }$. Introducing a new variable according
to $\sigma =\sigma ^{\prime }+\sigma _{0}$ and choosing $\sigma _{0}$ to
make the right hand side of eq. (\ref{hf}) of the first order in $C$, we put 
$\exp (-\varepsilon \sigma _{0})=CD$, where $D$ does not contain $C$. Then
after simple rearrangement we obtain $H=\frac{\alpha }{\delta }\sigma
+De^{-\delta \sigma }$ that agrees with (\ref{0111}). On the other hand,
there also exist solutions with $C\neq 0$ ( for instance, $[C,\Lambda ](0,0)$%
) which have no analogues among those with $C=0$.

Thus, we obtained the following qualitatively different cases.

Generic types: I$_{a}$, I$_{b}$, II$_{a}$, II$_{b}$, III.

Particular ones:

$[0,0](0,0);[0,0](\delta ,\alpha );$

$[0,\Lambda ](\delta ,\alpha );[0,\Lambda ](\delta ,0);[0,\Lambda ](0,\alpha
);$

$[C,\Lambda ](0,0);[C,\Lambda ](\delta ,0);[C,\Lambda ](\delta ,\alpha
);[C,0](\delta ,\alpha )$.

The case II$_{a}$ is equivalent to $[C,\Lambda ](\delta _{0},\alpha )$ and II%
$_{b}$ is equivalent to $[C,0](\delta _{0}$,$\alpha )$.

All other particular cases either are impossible or can be obtained directly
by letting the parameters their particular values.

\section{Examples}

In this section we restrict ourselves to examples that possess properties,
missed or overlooked in previously known exactly solvable models. Consider,
for example, the case $[0,\Lambda ](\delta ,0)$. It is convenient to
introduce a Schwarzschild coordinate $x$ according to $dx=d\sigma g$. Then
it follows from (\ref{0110}) that 
\begin{equation}
g=a\frac{H-H_{0}}{U}\text{,}
\end{equation}
\begin{equation}
\frac{dx}{d\phi }=B^{-1}\frac{H^{\prime }}{U}\text{,}  \label{xf}
\end{equation}
where the constant $B$ obeys the relationships $a=\Lambda /D$, $B=\delta /a$%
, $aB^{2}=1$.

If $H(\phi )=H_{0}$ at some $\phi =\phi _{0}$, we have a horizon. Meanwhile,
an additional horizon may appear at $H\rightarrow \infty $. As a result, we
may obtain black hole and cosmological horizons. Indeed, consider the model
for which at $\phi \rightarrow \infty $ 
\begin{equation}
U\backsim e^{\phi m}(1+U_{1}e^{-\phi })\text{, }H\backsim e^{\phi
n}(1+H_{1}e^{-\phi })\text{, }n,m>0\text{.}
\end{equation}
Then direct implication of eq. (\ref{0110}) shows that $g\backsim \exp [\phi
(n-m)]\backsim x-x_{h}$ ($x_{h}$ is the horizon value of $x$), 
\begin{equation}
R\backsim \exp [\phi (m-n-1)]\text{,}
\end{equation}
\begin{equation}
T_{1}^{1(PL)}\backsim (2n-m)\exp [\phi (m-n)]+const\exp [\phi (m-n-1)]+...%
\text{,}
\end{equation}

the Hawking temperature $T_{H}^{(1)}=\frac{\left| \delta \right| }{4\pi }$
at the black hole horizon at $\phi =\phi _{0}$ and 
\begin{equation}
T_{H}^{(2)}=\frac{\left| \delta \right| }{4\pi }\frac{(m-n)}{n}
\end{equation}
at the cosmological horizon $\phi =\infty $.

The value $\phi =\infty $ is indeed the horizon provided $n<m$, the
condition of regularity of the cosmological horizon reads $m\leq n+1$, the
finiteness of quantum stresses on this horizon occurs if $2n-m=0$. Al three
criteria are met for $m=2n$, $n\leq 1$, in which case $%
T_{H}^{(1)}=T_{H}^{(2)}$ and we obtain two horizons at thermal equilibrium,
quantum stresses being finite on them.

On the other hand, if $n<m\leq n+1$, $m\neq 2n$, the cosmological horizon is
regular but quantum stresses diverge on it.

One can observe that the solutions $[C,\Lambda ](0,0)$ and $[C,\Lambda
](\delta ,0)$ in the case $U=const$ ($\omega =0$) give the constant
curvature solutions $R=-2\sigma _{0}^{2}<0$ ( 2d adS metric)\ in the first
case and $R=e^{2H_{0}}DC\delta ^{2}$ in the second one (2D dS metirc, if $%
DC>0$). It was shown earlier that dS and adS metric appear in 2D dilaton
theories for constant dilaton solutions, $\left( \nabla \phi \right) ^{2}=0$ 
\cite{solod}, \cite{zasl98}. However, we see that the reverse is not
necessarily true: we obtained the constant curvature solutions with
essentially inhomogeneous dilaton field. This is due to $C\neq 0$, so these
solutions could not appear in previous studies of exactly solvable models 
\cite{exact}, \cite{thr}.

In the case [$C$, $\Lambda $]($0$, $0$) with $U=U_{0}=const$, the curvature $%
R=\frac{2U_{0}C}{1-2\kappa C}=const<0$ ($\Lambda $ and $C$ have different
signs according to (\ref{h00})) that is nothing else than the usual
two-dimensional adS space-time with an acceleration horizon. If the
potential is not constant identically but $U\rightarrow U_{0}=const$
asymptotically, we get a black hole extremal horizon.

If, say, at $\phi \rightarrow \infty $ $H\thicksim H_{1}\phi $ and $%
U=\Lambda e^{-\phi }$, we have for the same type of solutions $g\backsim
\sigma ^{m-2}$, $m=(CH_{1})^{-1}$. If $0<m<1$, $g\backsim (x-x_{h})^{n}$
with $n=\frac{2-m}{1-m}>2$. In this sense a horizon is ultraextremal.

$[0,0](0,0)$

Let at $\phi \rightarrow \infty $ $\eta \thicksim 2\phi $, $H\thicksim
e^{\phi }$. Then $g\thicksim \sigma ^{-2}$ and we again obtain an extremal
horizon. In fact, as in this example the potential $U=0$, we can choose the
function $\eta $ at our will. For instance, let $\eta =\ln ch\phi $, $%
H=sh\phi $. Then the solution is symmetric with respect to reflection $\phi
\rightarrow -\phi $, $g=\left( ch\phi \right) ^{-1}$ and at both infinities
we have extremal horizons in equilibrium. In so doing, they are
''ultracold'': $x\backsim \phi $ and $g\backsim \left( chx\right) ^{-1}$, so
not only the metric function but also their derivatives vanish at infinity.

Consider the case $H=e^{-2\phi }-\kappa \phi $, $\omega =-2$. It corresponds
to the solutions found in \cite{suk} in the cosmological context. It is
convenient to introduce the coordinate $\rho $ according to $d\rho =\sqrt{g}%
d\sigma $, the proper length $l=\left| \rho \right| $. Then after some
algebraic manipulations it follows from (\ref{000}) that $g=a^{2}$, 
\begin{equation}
a=\frac{\sqrt{b^{2}\rho ^{2}+2\kappa }+b\rho }{\kappa }\text{,}
\end{equation}
$b\,$is a constant. If $\kappa \rightarrow 0$, $b\rho <0$, we obtain $%
a\rightarrow \left| b\right| l^{-1}$. The region $b\rho >0$ does not have a
classical counterpart. The curvature 
\begin{equation}
R=-\frac{4\kappa b^{2}}{\left( b^{2}\rho ^{2}+2\kappa \right) ^{3/2}\left(
b\rho +\sqrt{b^{2}\rho ^{2}+2\kappa }\right) }
\end{equation}
is everywhere finite, including infinity. In the limit $\rho b\rightarrow
+\infty $ the metric function $g\backsim l^{2}$, the curvature $R\backsim
l^{-4}\rightarrow 0$. If $b\rho \rightarrow -\infty $, $g\backsim l^{-2}$, $%
R\backsim l^{-2}\rightarrow 0$. Thus, we have a nonextreme horizon at one
infinity and the Rindler metric at the other one.

The solution of $[0,0](\delta ,\alpha )$ type corresponds to the static
analogue of what is called ''the second branch'' in the cosmological context%
\cite{suk}.

It is worth noting correspondence between some types of exact solutions that
follows directly from the explicit formulas. For the solutions $[0,0](\delta
,\alpha )$ and $[C,\Lambda ](0,0)$ the dependence of the metric function on
dilaton $g(\phi )$ coincide provided $C=\delta ^{2}/2\alpha $; for the
solutions $[0,0](0,0)$ and $[C,\Lambda ](0,0)$ the spatial dependence of the
dilaton on the proper length coincide, provided the constant $\left|
A\right| =\left| C\sigma _{0}\right| $.

One can observe that the solutions $[C,\Lambda ](0,0)$ and $[C,\Lambda
](\delta ,0)$ in the case $U=const$ ($\omega =0$) give the constant
curvature solutions. It was shown earlier that dS and adS metric appear in
2d dilaton theories for constant dilaton solutions, $\left( \nabla \phi
\right) ^{2}=0$ \cite{solod}, \cite{zasl98}. However, we see that the
reverse is not necessarily true: we obtained the constant curvature
solutions with essentially inhomogeneous dilaton field. This is due to $%
C\neq 0$, so these solutions could not appear in previous studies of exactly
solvable models \cite{exact}, \cite{thr}.

\section{Summary}

Thus, we considered a rather wide family of exactly solvable models of 2D
dilaton gravity with backreaction of conformal fields, which includes
previously known particular models of this kind, and enumerated all possible
types of static solutions which appear in this family. In so doing, the
explicit results were listed for static solutions. However, if a time and
space variable are interchanged, we obtain (with signs properly reversed)
exact solutions for string-inspired cosmology that gives potential set for
the choice of everywhere regular space-times, detailed description of
inflation, etc. The list of solutions given above may also describe an
initial and final configurations in the problems of black hole formation and
evaporation. Further applications for black hole physics and cosmology
depend strongly on the concrete choice of the models.


\begin{references}
\bibitem{callan}  C. G. Callan, S. Giddings, J. A. Harvey, and A.
Strominger, Phys. Rev. D{\bf \ }45 (1992) R1005.

\bibitem{bil}  A. Bilal and C. G. Callan, Nucl. Phys. B 394, 73 (1993).

\bibitem{alw}  S. P. de Alwis, Phys. Rev. D 46, 5429 (1992).

\bibitem{rst}  J. G. Russo, L. Susskind, and L. Thorlacius, Phys. Rev. D{\bf %
\ }46 (1992) 3444; Phys. Rev. D{\bf \ }47 (1992) 533.

\bibitem{rob}  G. Michaud and R. C. Myers, Two-Dimensional Dilaton Black
Holes, gr-qc/9508063.

\bibitem{fub}  A. Fabbri and J. G. Russo, Phys. Rev. D 53, 6995 (1995).

\bibitem{kaz}  Y. Kazama, Y. Satoh, and A. Tsuichiya, Phys. Rev{\it .} D 51,
4265 (1995).

\bibitem{exact}  O. B. Zaslavskii, Phys. Rev. D 59, 084013 (1999).

\bibitem{thr}  O. B. Zaslavskii, Phys. Lett. B 459, 105 (1999).

\bibitem{gukov}  N. Berkovits, S. Gukov and B.C. Vallilo, Nucl.Phys. B 614,
195 (2001).

\bibitem{ext}  O.B. Zaslavskii, Phys. Lett. B 475, 33 (1999 ).

\bibitem{mod}  O. B. Zaslavskii, Mod. Phys. Lett. A 17, 1175 (2002).

\bibitem{extr2}  O. B. Zaslavskii, Exactly solvable models in 2D
semiclassical dilaton gravity and extremal black holes, hep-th/0211207 (To
appear in Class.Quant. Grav.).

\bibitem{dv}  D. Grumiller, W. Kummer and D. V. Vasilevich, Phys.Rept. 369
(2002) 327.

\bibitem{od}  S.Nojiri and S. Odintsov, Int. J. Mod. Phys. A 16, 1015 (2001).

\bibitem{fil1}  A. T. Fillipov and V. G. Ivanov, Phys.Atom.Nucl. 61, 1639
(1998) (hep-th/9803059).

\bibitem{fil2}  A. T. Fillipov, Mod.Phys.Lett. A 11, 1691 (1996).

\bibitem{eliz2}  E. Elizalde and S.D. Odintsov, Nucl. Phys. B 399, 581
(1993).

\bibitem{strobl}  T. Kloesch and T. Strobl, Class.Quant.Grav. 13, 965
(1996); Erratum-ibid. 14, 825 (1997).

\bibitem{pelzer}  H Pelzer and T. Strobl, Class.Quant.Grav. 15, 3803 (1998).

\bibitem{eliz4}  E. Elizalde, P. Fosalba-Vela, S. Naftulin, S. D. Odintsov,
Phys. Lett. B 352, 235 (1995).

\bibitem{pl}  A. M. Polyakov, Phys. Lett. B 103, 207 (1981).

\bibitem{cruz}  J. Cruz and J. Navarro-Salas, Phys. Lett. B 375, 47 (1996).

\bibitem{bose}  S. Bose, L. Parker, and Y. Peleg, Phys. Rev. D 52, 3512
(1995).

\bibitem{suk}  S. Bose and S. Kar, Phys. Rev. D 56, 4444 (1997).

\bibitem{solod}  S. N. Solodukhin, Phys. Rev. D{\bf \ }53, 824 (1996).

\bibitem{zasl98}  O. B. Zaslavskii, Phys. Lett.{\bf \ }B{\bf \ }424, 271
(1998).
\end{references}
\end{document}